\documentclass[a4paper, 11pt, onecolumn]{elsarticle}

\usepackage{lineno,hyperref}
\usepackage[utf8]{inputenc}
\usepackage{graphicx}
\usepackage{graphics}
\usepackage{float}
\usepackage{siunitx}
\usepackage{blindtext}
\usepackage[nice]{nicefrac}
\usepackage[dvipsnames]{xcolor}
\usepackage{amsmath}
\usepackage{lscape}

\graphicspath{{./figures/}}

\modulolinenumbers[5]

\journal{arXiv}

\begin{document}

\begin{frontmatter}

\title{Flight of a  honeybee in turbulent wind}

\author[1,2]{Bardia Hejazi\corref{mycorrespondingauthor}}

\author[1,2]{Christian K\"uchler}

\author[1,2]{Gholamhossein Bagheri}

\author[1,2,3,4]{Eberhard Bodenschatz\corref{mycorrespondingauthor}}
\cortext[mycorrespondingauthor]{Corresponding authors}
\ead{bardia.hejazi@ds.mpg.de (Bardia Hejazi), eberhard.bodenschatz@ds.mpg.de}

\address[1]{Laboratory for Fluid Physics, Pattern Formation and Biocomplexity, Max Planck Institute for Dynamics and Self-Organization (MPI-DS), 37077 G\"ottingen, Germany}
\address[2]{Max Planck University of Twente Center for Complex Fluid Dynamics, MPI-DS, 37077 G\"ottingen, Germany}
\address[3]{Institute for Dynamics of Complex Systems, University of G\"ottingen, 37077 G\"ottingen, Germany}
\address[4]{Laboratory of Atomic and Solid State Physics and Sibley School of Mechanical and Aerospace Engineering, Cornell University, 14853 Ithaca, NY, USA}

\begin{abstract}

In windy conditions, the air is turbulent. The strong and intermittent velocity variations of turbulence are invisible to flying animals. Nevertheless, flying animals, not much larger than the smallest scales of turbulence, manage to maneuver these highly fluctuating conditions quite well.  Here we quantify honeybee flight with time-resolved three-dimensional tracking in calm conditions and controlled turbulent winds. We find that honeybee mean speed and acceleration are only weakly correlated with the strength of turbulence. In flight, honeybees accelerate slowly and decelerate rapidly, i.e., they break suddenly during turns and then accelerate again.  While this behavior is observed in both calm and turbulent conditions, it is increasingly dominant under turbulent conditions where short straight trajectories are broken by turns and increased maneuvering. This flight-crash behavior is reminiscent of turbulence itself. Our observations may help the development of flight strategies for miniature flying robotics under turbulent conditions.

\end{abstract}


\end{frontmatter}

\section{Introduction}

How animals manage to navigate complex turbulent flows is important and of interest to biologists, physicists, engineers, and also the general public. For example, mosquito's prefer not to fly in turbulent conditions~\cite{Hoffmann2002}, possibly associated with the high accelerations in the flow \cite{Voth2002,LaPorta2001}.  Recent studies have investigated the behavior of small organisms such as planktonic gastropod larvae in turbulence \cite{DiBenedetto2022} or of  larger birds  such as a golden eagle in atmospheric turbulence~\cite{Laurent2021}. The ability for animals and especially insects to navigate turbulent flows while in flight has been studied to determine how they are capable of maintaining flight stability and control~\cite{Sun2014}. Such studies can further improve small-scale robotics for visually guided flight~\cite{Srinivasan2011}, energy harvesting strategies \cite{bollt_bewley_2021} and enhance flapping wing designs~\cite{Helbling2018}.

The study of insect flight has gathered much interest due to their relative ease of handling and performing controlled laboratory experiments in different environments, such as the study of fruit flies~\cite{Fuller2014,Ortega2018} and midges~\cite{Kelly2013}. Particular insects of interest have been the honeybee, bumblebee  and orchid bee since they are vital to the environment as natural pollinators~\cite{Hung2018,Delaplane2000}. The majority of experiments studying bee flight are conducted in laboratory settings with field experiments mostly focusing on bee hive entrance activity~\cite{Peters2019,Butler1952}, and monitoring for counting statistics~\cite{Magnier2018,Thi2019,Chiron2013,Chiron2013II,Babic2016}. Most experiments performed in the laboratory examine bee flight (bumblebees, orchid bees, etc) in laminar flow and study wing aerodynamics and motion~\cite{Altshuler2005}. Other experiments have investigated how bee flight and behavior changes due to external factors such as the presence of predators~\cite{Lenz2012}. Experimental studies examining the influence of turbulent flow in laboratory settings investigated changes in flight in the form of how wing dynamics and bee orientations changed while flying in different conditions~\cite{Crall2017,Jakobi2018,Ravi2013}. Studies have shown that bees have increased rolling instabilities in turbulent conditions~\cite{Crall2017,Ravi2013}, with field measurements also showing stability enhancing behavior to improve rolling stability in the form of bees extending their hind legs during flight~\cite{Stacey2009}. Numerical simulations have shown that there are no significant changes in cycle-averaged aerodynamic forces, moments, and flight power in strong turbulent flow conditions  as compared to flight in laminar flow. However, the variance of aerodynamic measures increases with turbulence intensity~\cite{Engels2016}.

The natural flight environment of honeybees are turbulent winds with structures spanning a wide range of scales. Atmospheric turbulence consists of time scale of less than 1 second to typically 1 hour and the corresponding length scales are from millimeter up to the boundary layer thickness of hundreds of meters~\cite{Wyngaard1992,Holtslag2003}.
In this study we report data on  the flight of honeybees in their natural habitat while flying towards and away from the hive.  We reconstruct honeybee trajectories and study honeybee flight paths in natural windless conditions and in the presence of artificially generated turbulent wind using a fan and active grid system.


\section{Methods}
 \label{sec:Methods}


\begin{figure*}[ht]%
\centering
\includegraphics[width=1\textwidth]{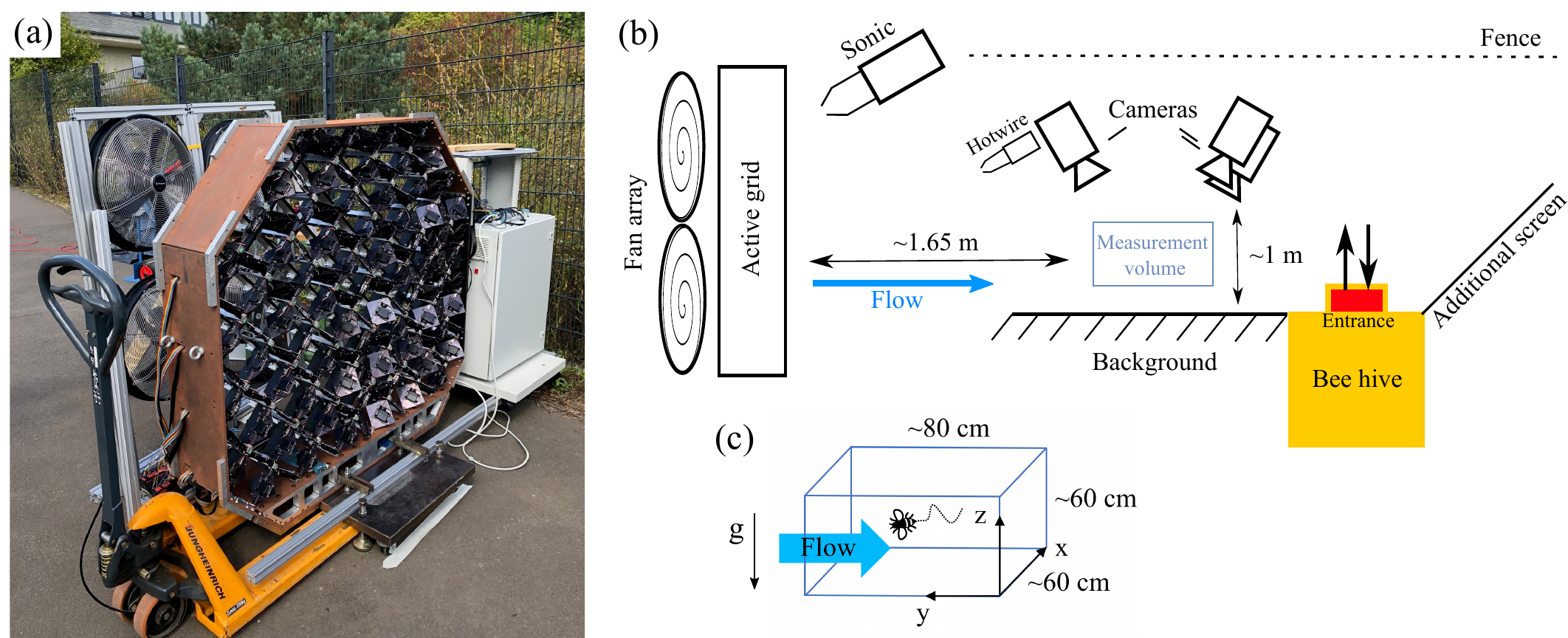}
\caption{(a) Active grid used to generate turbulent flow with different characteristics. (b) Schematic of the experimental setup showing the location of the honeybee hive, cameras, fans and active grid, and flow measurement devices. The fence and additional screen limit available routes and encourage the honeybees to fly though the detection volume. The fence and vegetation is also visible behind the fan and active grid in (a). (c) The approximate dimensions of the measurement volume where camera images overlap.}
\label{setup}
\end{figure*}

Experiments were performed at the Max Planck institute for Dynamics and Self-Organization (MPI-DS) in Göttingen, Germany.
Experiments were carried out between September 7-13, 2021, between the hours of 11:00-15:00 each day. The weather during the experiments was calm with little to no natural wind and no precipitation during experiments.
As shown in Figure~\ref{setup} experiments were performed near the entrance of a hive with the honeybee species \textit{Apis mellifera carnica}. Our institute at the time of experiments was in possession of three hives and experiments were performed near the entrance of one of these hives.

\subsection{Turbulence generation and characterization}
Wind and turbulence was produced using a fan array and active grid system identical to the one used in wind-tunnel experiments at the variable density wind-tunnel in Göttingen~\cite{Bodenschatz2014,Kuchler2019}. The fan array consisted of 4 fans with three adjustable speed setting that blew air into to the active grid. The active grid was controlled with an in-house program capable of moving each grid flap independently allowing for adjustable turbulence length and time scales, more details on the controlling code and active grid can be found in~\cite{Griffin2019}.
Values for few of the flow quantities measured are given in Table~\ref{table} along with the number of tracks in those conditions.

To identify the characteristics of the turbulent flow we used a thermal anemometer system previously used in high Reynolds number flows~\cite{Kuchler2019}. We also used a sonic anemometer used in atmospheric measurements~\cite{Schroder2021} to measure 3D wind speeds. Here we estimate the turbulence intensity and characteristics using data from the thermal anemometer.

The sonic anemometer was used to determine wind direction and speed for when fans were active and for when fans were off~\cite{SM2022}. The sonic anemometer had a sampling frequency of 30 Hz. From the data we clearly see the times in which the fans were turned on and off. When fans were on, the wind direction was strongly in the same direction as where the fans were blowing. When fans were off, there was no clear direction to the wind and the mean wind speed was 15\% of the mean wind speed when the fans were on. This shows that the weather during the experiments was mild with no strong natural winds present.
We also checked wind speeds recorded by a cup anemometer on a nearby building at our institute. The cup anemometer also recorded low wind speeds on the days of experiments with slightly higher values recorded than the sonic anemometer. This is most likely due to the fact that the cup anemometer was on the roof of a building and recorded slightly higher wind speeds because of it.

\begin{table}[ht]
\footnotesize
\begin{center}
\caption{Turbulence characteristics for experiments along with the number of tracks in each experiment. Here $Re_{\lambda}=u' \lambda/\nu$ is the Taylor microscale Reynolds number, where $u'$ is the root mean square fluctuating velocity, $\lambda=\sqrt{15\nu/\varepsilon}u'$ is the Taylor microscale, $\nu$ is the kinematic viscosity of the fluid ($1.5\times 10^{5}~\text{m}^{2}/\text{s}$ for air), and $\varepsilon$ is the turbulent energy dissipation rate. The mean flow velocity is given by $U$. The Kolmogorov timescale is $\tau=(\nu/\varepsilon)^{1/2}$ and the Kolmogorov lengthscale is $\eta=(\nu^{3}/\varepsilon)^{1/4}$.
Experiment 0 is for when fans were off and experiments 1-8 are for when fans were on and wind was generated.
A higher number of tracks were recorded for the first day of experiments on 09/07/2021, Exp. 1 and 2, due to movement and experiment setup around the hive which caused agitation and higher than usual activity at the hive. For the remaining days the honeybees were less agitated and had adjusted to the changes and activity near the hive.
}\label{table}%
\begin{tabular}{@{}lllllllll@{}}
\hline
Ex. & $R_{\lambda}$  & $\eta$ (mm) & $\tau$ (ms) & $\varepsilon$ ($\text{m}^2/\text{s}^3$) & $u'$ (m/s)& $U$ (m/s) & Tracks & Date (mm/dd/yyyy) \\
\hline
1    & 389  & 0.26 & 4.3 & 0.79 & 0.60 & 2.96 & 332 & 09/07/2021\\
2    & 336  & 0.28 & 5.1 & 0.57 & 0.51 & 2.66 & 214 & 09/07/2021\\
3    & 282  & 0.27 & 4.6 & 0.70 & 0.49 & 2.50 & 58 & 09/08/2021\\
4    & 260  & 0.27 & 4.8 & 0.64 & 0.46 & 2.16 & 71 & 09/08/2021\\
5    & 333  & 0.26 & 4.3 & 0.80 & 0.55 & 2.31 & 47 & 09/08/2021\\
6    & 266  & 0.28 & 5.2 & 0.56 & 0.45 & 2.37 & 59 & 09/09/2021\\
7    & 265  & 0.30 & 5.8 & 0.45 & 0.42 & 2.22 & 51 & 09/09/2021\\
8    & 239  & 0.29 & 5.3 & 0.53 & 0.42 & 1.84 & 41 & 09/09/2021\\
0    & -  & -  & -  & - & - & - & 88 & 09/13/2021\\
\hline
\end{tabular}
\end{center}
\end{table}

\subsection{Imaging and tracking}
Imaging was done using three GoPro Hero 9 Black cameras recording at 120 frames per second at a resolution of $2704 \times 1520$. 
Using GoPro cameras is a relatively simple and cost effective method of performing biological experiments in the field~\cite{Jackson2016}. The cameras were on the linear setting which digitally corrects for optical distortions caused by the lens with internal software. Calibrations were performed using a $51 \times 31$ grid of circular black dots on a white background that almost spanned the complete imaging volume. The dots were 0.5 cm in diameter and 1 cm apart from each other.
A blank white background was used for shadow imaging of the honeybees similar to methods previously used capable of highly precise imaging and tracking of particles in turbulence~\cite{Hejazi2019,Hejazi2021}. The three cameras had different viewing angles with an overlap region which created a measurement volume of approximately $80 \times 60 \times 60$ cm in size. The existing fencing and an additional screen, not within view of the cameras, encourage the honeybees to fly through the measurement volume (see Figure~\ref{setup}).

Cameras are triggered by a remote able to control all three cameras simultaneously. 
Each experiment with specific flow characteristics defined by active grid settings consists of 6, 5 minute long videos, totalling to 30 minutes of data for each experiment.
The limiting factor for imaging with the GoPro cameras used here is firstly the battery life which allows for approximately 1 hour of recording. After the batteries are drained, a replacement set is used and a new set of calibration images are acquired. The second limitation is the overheating of the cameras which occurs due to continuous recording. During wind experiments this issue is resolved due to the cooling nature of the blowing wind. However during no wind experiments cameras switch off after overheating and a waiting period is required before resuming experiments. After the waiting period is done and cameras are turned back on, a new set of calibration images are also acquired to account for any changes in camera position during the pressing of the on button located on the camera bodies.




\begin{figure*}[ht]%
\centering
\includegraphics[width=1\textwidth]{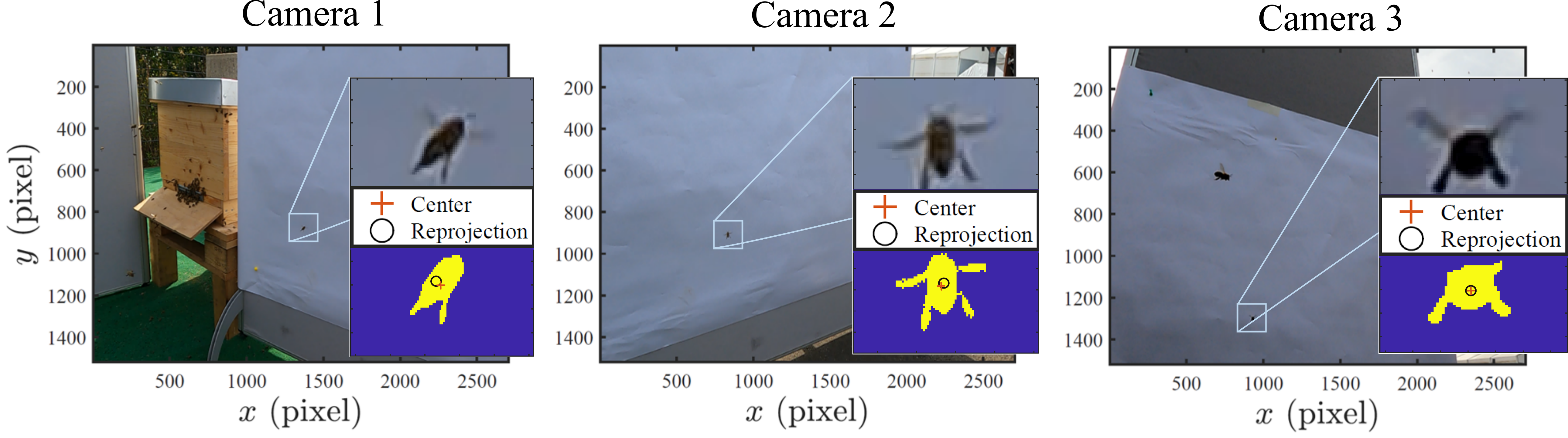}
\caption{Images from the three GoPro cameras used in our experiments. The raw image as seen by the three views and a zoomed in picture of a detected honeybee with its binarized image shown below. The center of the honeybee is found using the center of mass of the bright pixels (red crosses) which is then used for steromatching and finding the 3D coordinates of the honeybee. The reprojection from the stereomatched 3D position back to the 2D camera plane is also shown to demonstrate the tracking quality (black circles).}
\label{imaging}
\end{figure*}

Identifying honeybees from the images and obtaining tracks was done with in-house codes based on previously used methods~\cite{Hejazi2019,Hejazi2021}. The raw images were cropped to only include sections with the white background. The images are then background subtracted, inverted, and binarized with an appropriate threshold. The honeybees now appear as bright pixels with a dark background where the honeybee positions are then identified using the center of mass of the bright pixels. The positions of the honeybees from the three cameras are then used for steromatching to obtain 3D (three-dimensional) positions and tracks.

Camera syncing is checked by calculating reprojection errors for the three cameras. 
To estimate tracking error the 2D (two-dimensional) positions are stereomatched to obtain 3D positions and the 3D positions are again converted to their original 2D image plane positions to obtain a reprojection error, defined as the mean distance between the position of original centers found and reprojected centers across all three cameras. 
The frames are then adjusted such that the reprojection error is minimized and we achieve optimal camera syncing not achievable solely with the remote triggering.
Furthermore, we discard any points with reprojection errors larger than 10 pixels as bad matches. In this way we are able to reduce tracking errors where for all available data, a mean imaging reprojection error of 1.6 pixels is achieved. This error is an acceptable value since each honeybee is approximately $\sim 10^3$ pixels in size in each image and the reprojection error in finding honeybee positions can be attributed to the center of mass position identification method where the positions identified in the 2D image of the three cameras may not be exactly the same position in real-life coordinates. Additional errors may be caused by the cameras still being slightly out of sync even after optimization.
The imaging and reprojection for an example frame is shown in Figure~\ref{imaging} after optimization.
Future work can improve imaging by fitting a model honeybee to the images to find the same center on each camera, which would require us to use higher resolution cameras capable of exact syncing.

After obtaining honeybee 3D positions, honeybee velocities $\vec{v}$ and accelerations $\vec{a}$ are obtained by fitting a second order polynomial to the 3D position data. The polynomial is of the form $p(x)=p_{1}x^{2}+p_{2}x+p_{3}$ where, $p_{2}=v$ and $p_{1}=a$. Here we only consider tracks for further analysis that are longer than 15 frames. The fit-length used for fitting polynomials is $h=7$ where a track with length $l$ is split into $l-2h$ sections for polynomial fitting.
The rest of the analysis follows from these values.

\section{Results}

\begin{figure*}[ht]%
\centering
\includegraphics[width=1\textwidth]{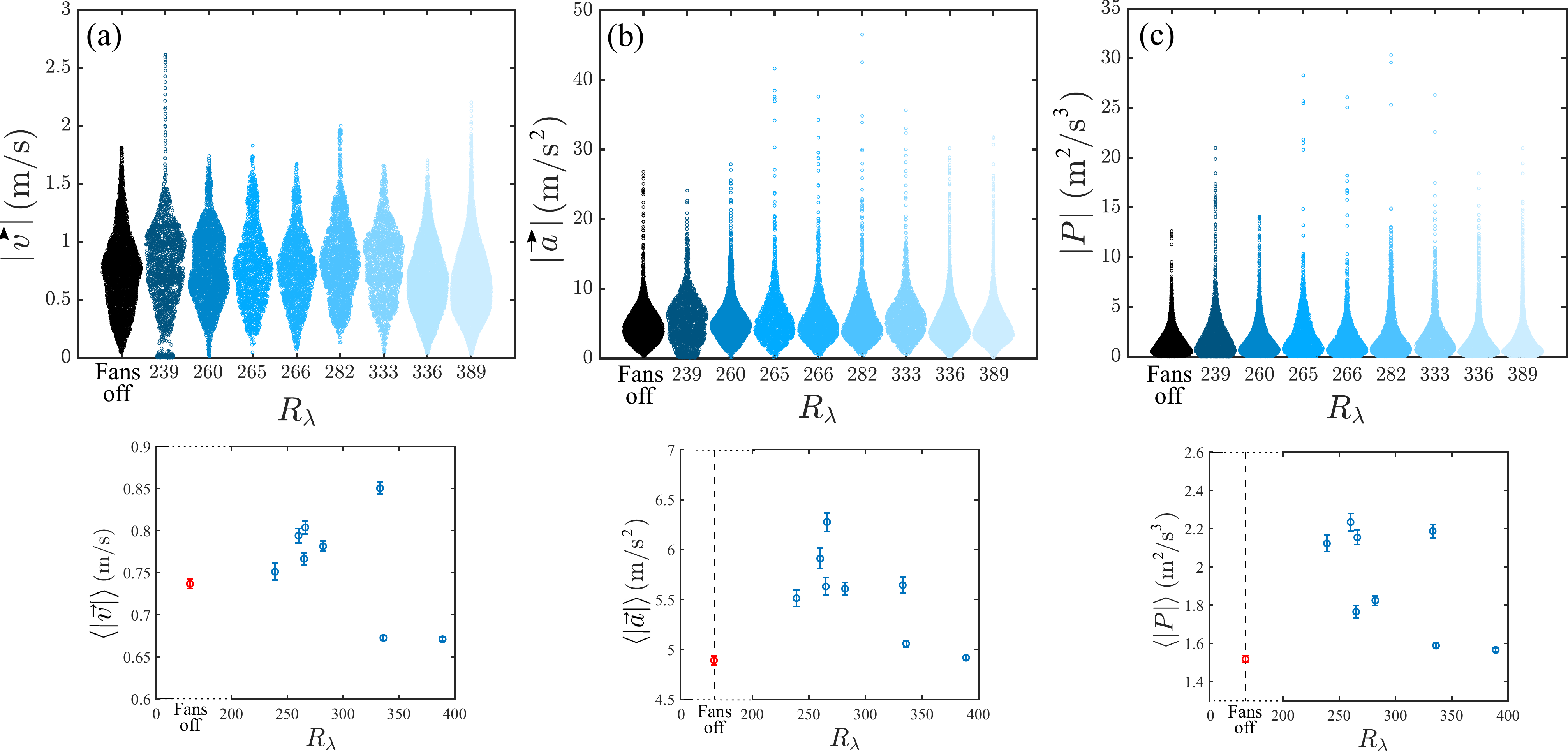}
\caption{(a-c) The magnitude of the measured values for honeybee velocity, acceleration, and flight power ($P=\vec{a} \cdot \vec{v}$) in windless conditions ($R_{\lambda} \ll 100$, fans off) and at different turbulence levels (fans and active grid on). The values for $v$, $a$, and $P$ appear to be uncorrelated with the turbulent intensity $R_{\lambda}$. Here $Re_{\lambda}=u' \lambda/\nu$ is the Taylor microscale Reynolds number, where $u'$ is the root mean square fluctuating velocity, $\lambda$ is the Taylor microscale, and $\nu$ is the fluid kinematic viscosity. Smaller sub plots are the mean values in each experiment as a function of $R_{\lambda}$, showing no immediate trend.}
\label{violin_mean}
\end{figure*}

Figure~\ref{violin_mean} (a-c) shows the magnitude of the velocity, acceleration, and flight power, of honeybees as violin plots for when fans are off and windless is generated and when fans are on and generating turbulent wind with the active grid. The flight power per insect mass is calculated using a formulation know from Lagrangian turbulence to define the rate of change of the kinetic energy following a tracer particle~\cite{Xu2014}, where the power per mass, $P=\vec{a} \cdot \vec{v}$. Here we assume a negligible variance in the mass between individual honeybees compared to the large variance in acceleration and velocity. We generated wind with different turbulence levels represented by the Taylor microscale Reynolds number $R_{\lambda}\sim \sqrt{Re}$. 
No clear trend relating the mean values of $v$, $a$, and $P$ to turbulence intensity can be discerned in the smaller sub-plots, or from the shape of the probability distributions shown in the violin plots.  

This is in agreement with previous numerical studies of bee flight where aerodynamic quantities averaged over wing-beat cycles do not change significantly in strong turbulence~\cite{Engels2016}, and field experiments showing that mean bee foraging activity is not influenced by windy conditions~\cite{Crall2017}. 
In our experiments, the mean values for the honeybee across all datasets are $\langle \lvert \vec{v} \rvert \rangle = 0.71$~m/s, $\langle \lvert \vec{a} \rvert \rangle = 5.18~\text{m}/\text{s}^2$, and $\langle \lvert  P \rvert \rangle = 1.70~\text{m}^2/\text{s}^3$. 

From the data obtained using the sonic anemometer we have measured the mean wind velocity and direction when fans are off and for when fans are active~\cite{SM2022}. The mean wind direction while fans are active is clearly in the direction which fans are blowing and the mean wind velocity is between 1.8 and 3 m/s for all  experiments. In contrast, when fans are off all three components of the wind velocity behave similarly with mean wind speeds of approximately 0.3 m/s. This clearly shows the difference in conditions when fans are active and wind is generated and when fans are off and no generated wind is present.
To better determine whether the strong directional generated wind has any effect on honeybee flight, we have combined all data with fans active into one data-set and labeled it as all wind data. We can then compare this dataset with data that was obtained with no fans on, labeled no wind data.

\begin{figure}[ht]%
\centering
\includegraphics[width=0.6\textwidth]{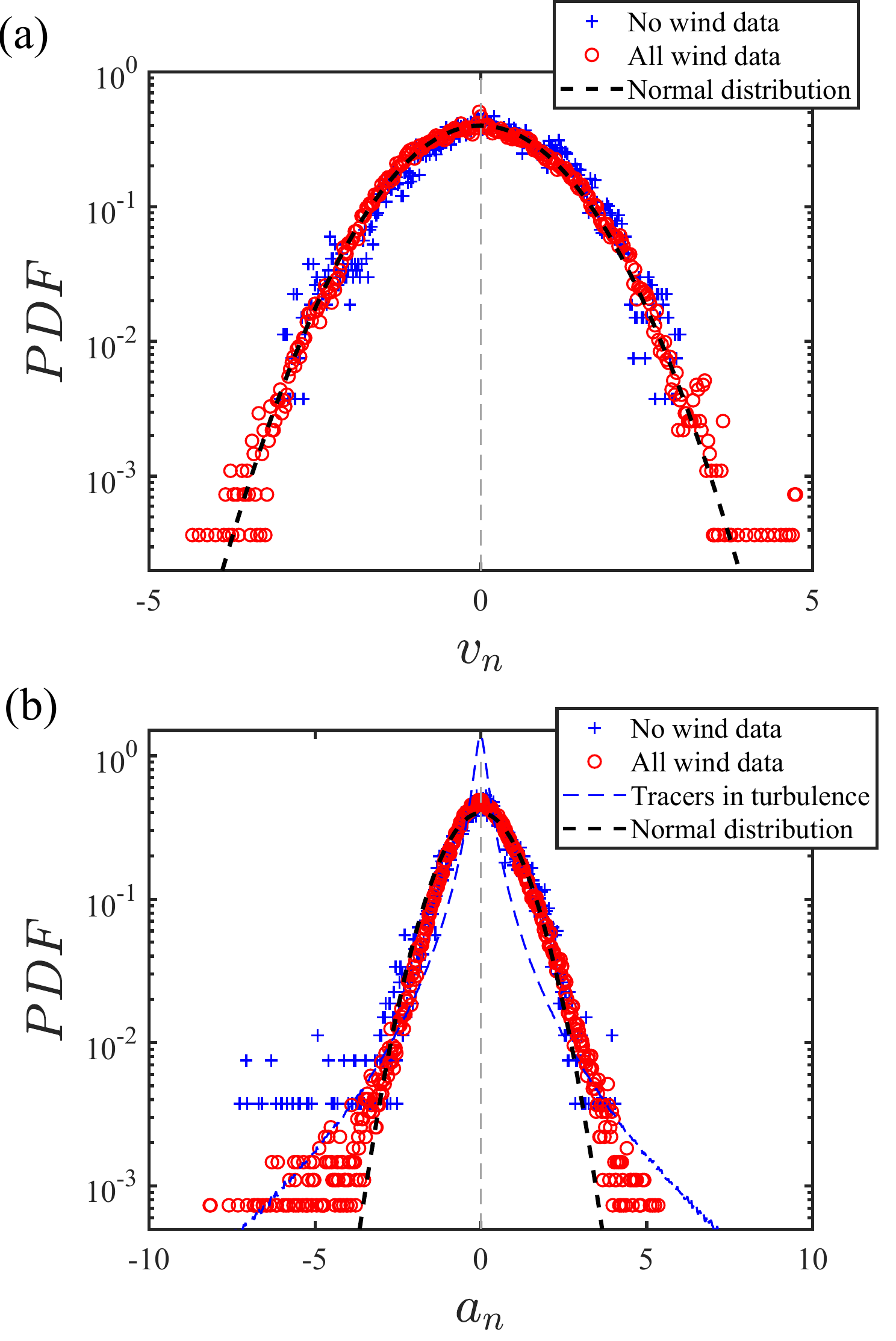}
\caption{(a) The probability distribution function (PDF) of honeybee velocities and (b) accelerations. Honeybees show control in turbulence where the acceleration PDF closely follows a normal distribution, however signs of turbulence become visible in the longer tail of the PDF. Here, the velocity PDF is defined as $P(v_{n})=(1/3)P(v_{i}/v'_{i})$ where $v'_{i}$ is the standard deviation of $v_{i}$, $a_{n}$ is similarly defined.}
\label{pdfs}
\end{figure}

Figure~\ref{pdfs} (a) and (b) show the probability distribution function (PDF) of honeybee velocities and accelerations.  The honeybee velocity follows a normal distribution as expected.
The closely normal distribution of the acceleration PDF shows that honeybees are able to maintain control while flying in turbulent conditions. Honeybees are active flyers since they are continuously flapping their wings during flight as compared to a passive flyer such as the soaring golden eagle that uses turbulence to its advantage and has acceleration PDFs more closely resembling tracers in turbulence~\cite{Laurent2021}, with much longer tails as compared to the honeybee. 
However, the accelerations while mostly close to a normal distribution, show signs of influence from the turbulent flow in the long tail of the acceleration PDF similar to particles in turbulence that reveal the intermittent nature of turbulence~\cite{LaPorta2001,Voth2002}. 

\begin{figure}[ht]%
\centering
\includegraphics[width=0.5\textwidth]{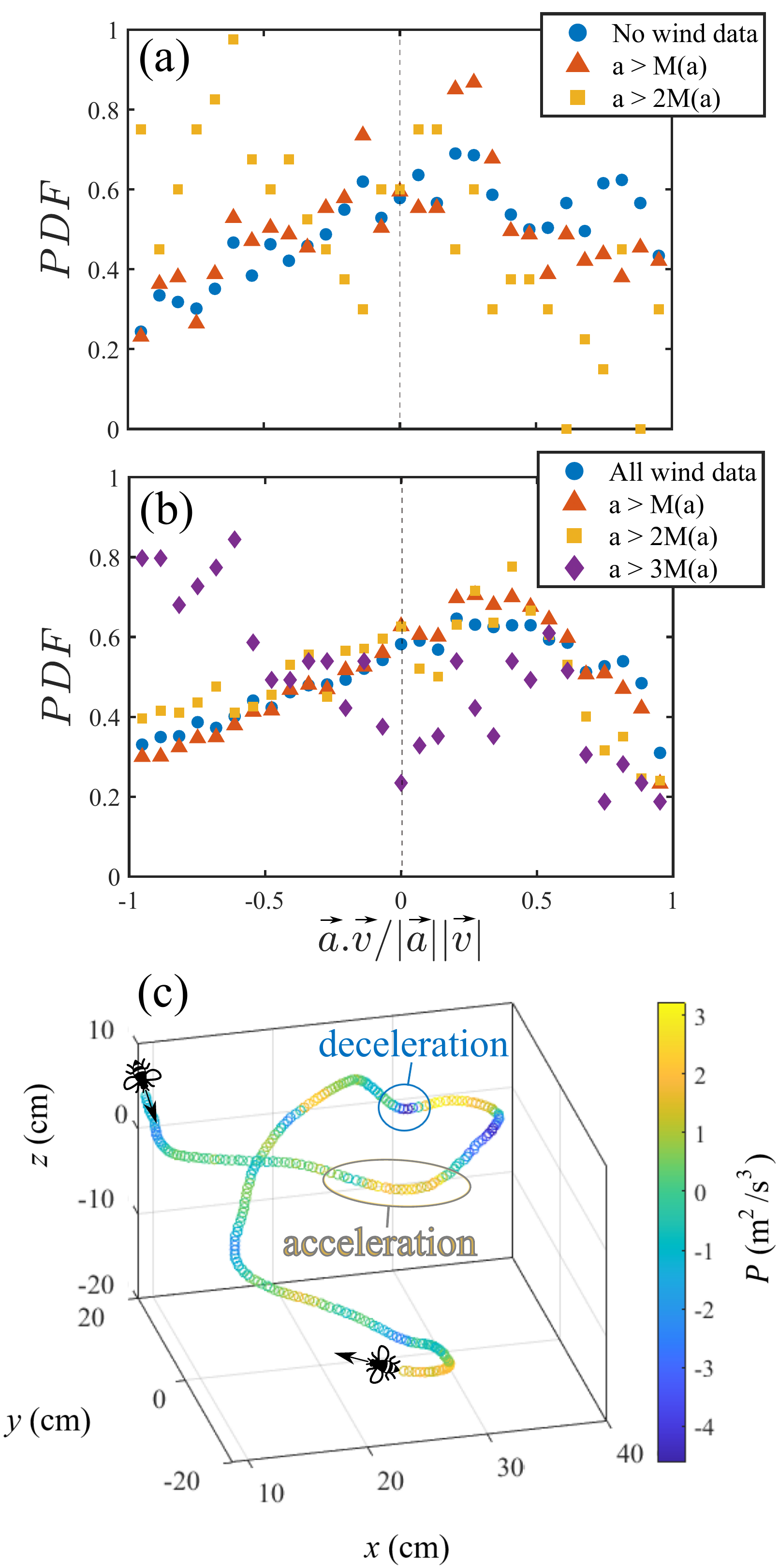}
\caption{(a-b) PDF of the cosine of the angle between $\vec{a}$ and $\vec{v}$, $\vec{a}\cdot \vec{v}/ \lvert \vec{a} \rvert \lvert \vec{v} \rvert$, for honeybee flight in (a) no wind conditions and (b) all wind conditions. The cosine is also conditioned on acceleration. The honeybee spends more time accelerating than decelerating where slow down events are sudden and have large negative accelerations. 
(c) Example of a honeybee trajectory showing the long time it takes for a honeybee to accelerate and the sudden rapid deceleration when slowing down where color shows the gain and loss in flight power.}
\label{3dtrack}
\end{figure}

To further understand the flight dynamics, we investigate the cosine of the angle between $\vec{a}$ and $\vec{v}$. A small angle between $\vec{a}$ and $\vec{v}$ indicates that the honeybee is acting to speed up the motion in the current direction of flight, whereas a larger angle indicates a change in flight direction. Figure~\ref{3dtrack} shows $\vec{a}\cdot \vec{v}/ \lvert \vec{a} \rvert \lvert \vec{v} \rvert$ for all the data with no wind in (a) and all wind data in (b). From these PDFs we see that the honeybees spend the majority of their flight time in an accelerating mode, meaning that the honeybees require more time to accelerate to a certain speed, while they can rapidly decelerate and lower their velocity. 
This is further demonstrated by conditioning $\vec{a}\cdot \vec{v}/ \lvert \vec{a} \rvert \lvert \vec{v} \rvert$ on the honeybee accelerations where we see that for extreme values of the acceleration, $a>2\text{M}(a)$ for no wind and $a>3\text{M}(a)$ for with wind, this value is more likely negative and that the honeybee is mostly decelerating. Here M stands for the median and $a>3\text{M}(a)$ data is not shown for the no wind data due to less available statistics in no wind conditions, however even with $a>2\text{M}(a)$ the increase in deceleration likelihood is observed. This lower acceleration threshold for observing deceleration's in no wind conditions can be due to the lower power needed for the honeybee to decelerate with no wind present. This conditioning shows that deceleration events are sudden and result in large values of the acceleration.


An example of a honeybee track is shown in Figure~\ref{3dtrack}~(c) where the slow accelerating and sudden decelerating regions are shown. Here the color shows the kinematic power of the honeybee flight. The accelerating section of the trajectory is a region where flight power slowly builds up and the honeybee increases its velocity. The decelerating section involves an abrupt turn where the power suddenly drops and the honeybee reduces its velocity as the turn is initiated followed by a power gain as the honeybee builds-up velocity once more. This phenomenon is a well-known feature for neutrally buoyant tracer particles in a turbulent flow, which experience so-called \textit{flight-crash} events of strong intermittent power loss~\cite{Xu2014}. For the case of a honeybee crash events occur when abrupt turns are made.

\begin{figure}[ht]%
\centering
\includegraphics[width=0.75\textwidth]{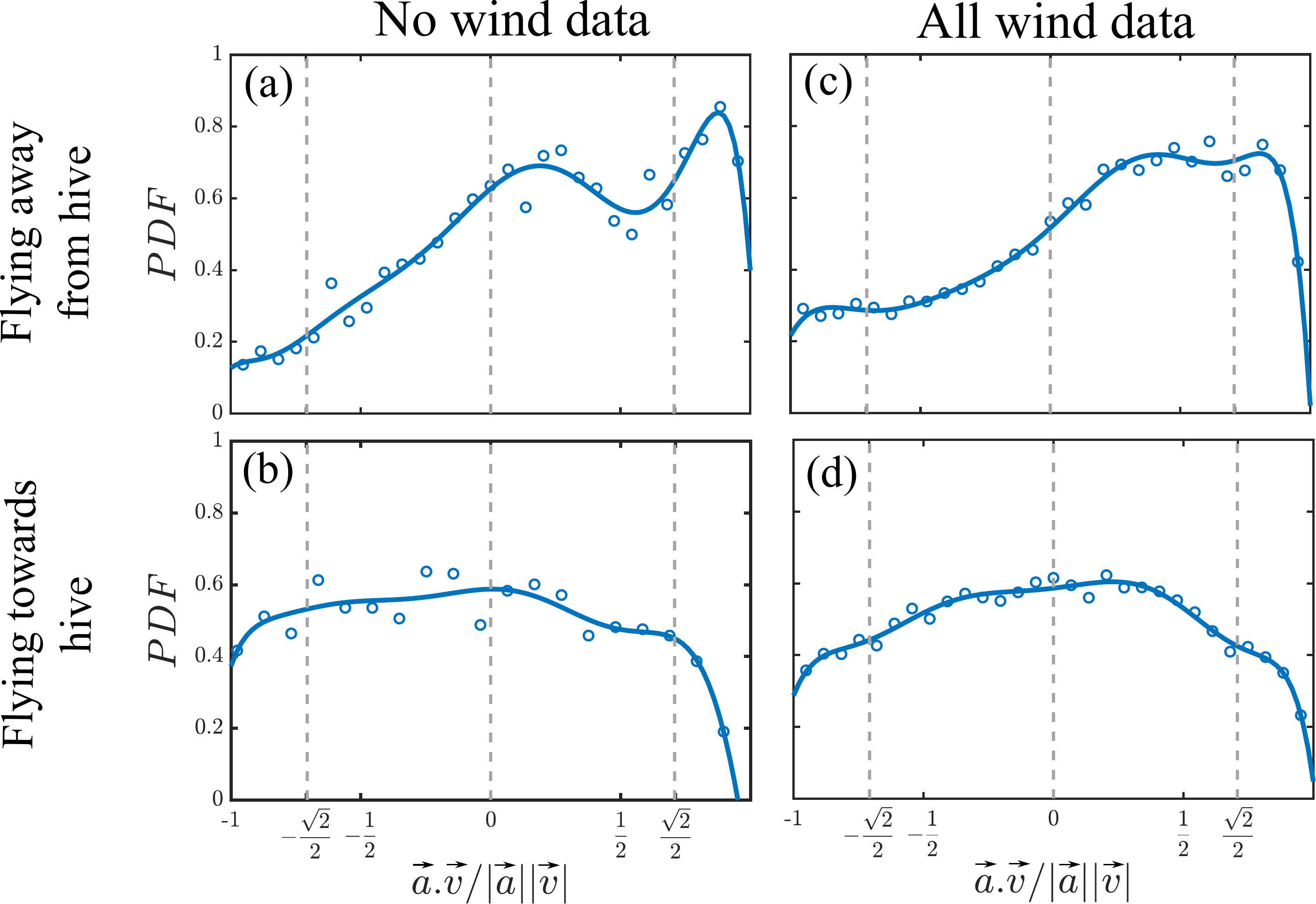}
\caption{The PDF of the cosine of the angle between $\vec{a}$ and $\vec{v}$ of honeybee flight for no wind conditions while flying away from the hive (a) and while flying towards the hive (b). Also shown for windy conditions while flying away from the hive or flying against the wind (c) and flying towards the hive or flying in the same direction as the wind (d). In no wind conditions, while flying away from the hive, the honeybee flies straight and makes abrupt turns. While flight in windy conditions mostly involves turning and evasive maneuvers, while more straight flight is suppressed. The dashed gray lines at $\pm \sqrt{2}/2$ show when the translational and centripetal accelerations are equal $\lvert \vec{a}_{t} \rvert = \lvert \vec{a}_{c} \rvert$. The solid blue lines are for better visualization and are polynomial fits to the data shown with circles.}
\label{away and towards no cond}
\end{figure}

We further examine honeybee flight by splitting tracks into flight away and towards the hive. This allows us to better analyze the influence of head-on flight against the wind (head wind) and flight in the same direction as the wind (tail wind).
We determine flight direction using the direction of the honeybee velocity vector along the $y$ coordinate axis. $y$ is both the direction of travel away and towards the hive and the direction of the wind mean flow.
Figure~\ref{away and towards no cond} shows $\vec{a}\cdot \vec{v}/ \lvert \vec{a} \rvert \lvert \vec{v} \rvert$ for the split tracks for no wind data flying away from the hive (a) and towards the hive (b), and for all wind data flying away from the hive or flying against the wind (c) and flying towards the hive or flying in the same direction as the wind (d).


Interestingly, the honeybee is increasing its overall cruising speed ($\vec{a}\cdot \vec{v}>0$) more frequently than decreasing it ($\vec{a}\cdot \vec{v}<0$) as it leaves and flies away from the hive in both no wind and windy conditions. And for times when flying towards the hive, the accelerating and decelerating events are more evenly distributed throughout its flight, for both no wind and windy conditions, signifying its anticipation of landing back at the hive and the need to reduce velocity.
Of particular interest is the comparison between Figure~\ref{away and towards no cond} (a) and (b), here the honeybee is mostly accelerating, however, in no wind conditions we clearly see two maxima when $\vec{a}\cdot \vec{v}/ \lvert \vec{a} \rvert \lvert \vec{v} \rvert>0$.
This indicates that when wind is not present the honeybee flies straight and makes abrupt turns further demonstrating its flight-crash like behavior~\cite{Xu2014}. Straight flight is shown by the maxima at $\vec{a}\cdot \vec{v}/ \lvert \vec{a} \rvert \lvert \vec{v} \rvert>\sqrt{2}/2$, and the abrupt turns are shown by the maxima at $0<\vec{a}\cdot \vec{v}/ \lvert \vec{a} \rvert \lvert \vec{v} \rvert<\sqrt{2}/2$.
Here the dashed lines at $\pm \sqrt{2}/2$ show regions where the translational acceleration $a_{t}$ and centripetal acceleration $a_{c}$ are equal, where $\vec{a}=\vec{a}_{t}+\vec{a}_{c}$.
In the case of Figure~\ref{away and towards no cond} (a), when $\vec{a}\cdot \vec{v}/ \lvert \vec{a} \rvert \lvert \vec{v} \rvert>\sqrt{2}/2$, the honeybee acceleration is predominantly in the form of translational acceleration $a_{t}>a_{c}$, which signifies straight flight. 
While the second maxima at $0<\vec{a}\cdot \vec{v}/ \lvert \vec{a} \rvert \lvert \vec{v} \rvert<\sqrt{2}/2$ shows that the honeybee is making turns and most acceleration is in the form of centripetal acceleration $a_{c}>a_{t}$. 
However, in windy conditions the honeybee no longer shows two distinct maxima while flying away from the hive as seen by Figure~\ref{away and towards no cond} (c) and the flight is spread across a range of $\vec{a}\cdot \vec{v}/ \lvert \vec{a} \rvert \lvert \vec{v} \rvert>0$. This shows that the honeybee is flying with a range of different centripetal accelerations, where the flight trajectory is perturbed and the bee is twisting and turning with increased crash events as it flies against the wind, while translational acceleration and more straight flight is suppressed.

To quantify the increase in maneuverability of the honeybee in turbulent conditions we use the trajectory curvature $k$ used to quantify the curvature of Lagrangian trajectories in turbulence~\cite{Xu2007}. The curvature has also previously been used to quantify the maneuverability of midges~\cite{Puckett2014}. The curvature of a trajectory is defined as $k=a_{n}/\lvert \vec{v} \rvert^2$, where $a_{n}=\lvert \vec{a} \times \vec{v} \rvert/\lvert \vec{v} \rvert$ is the magnitude of the normal acceleration to the flight direction. The value for the curvature is large whenever abrupt turns are performed and such events are distinct from the background~\cite{Puckett2014}. Similar to methods used by Puckett \textit{et al.}~\cite{Puckett2014} to define significant changes in motion, we define significant changes to be instances when the curvature exceeds the mean curvature for a particular experiment. Here, the mean curvature in calm conditions is 0.025~mm$^{-1}$ and for all wind data the mean curvature is 0.022~mm$^{-1}$. 
To determine the maneuverability change between calm and turbulent conditions, we calculate the total length of high curvature sections of trajectories in a given dataset and normalize by the total length of that dataset. In this way, we find that in calm conditions the normalized high curvature sections of motion is 0.1483 and in turbulent conditions this value is 0.1914, clearly showing the increase in honeybee maneuverability in turbulent conditions.

\begin{figure}[ht]%
\centering
\includegraphics[width=0.75\textwidth]{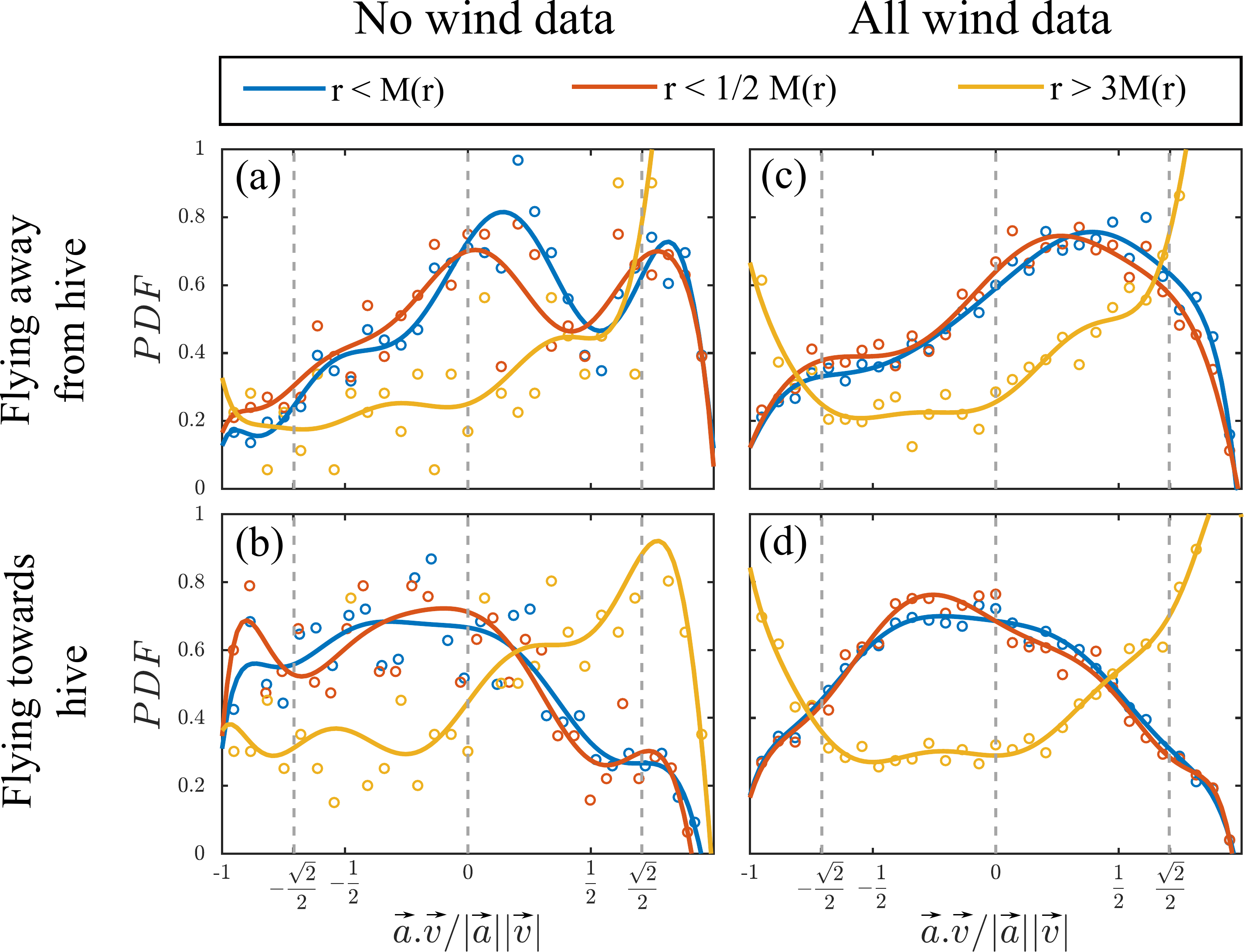}
\caption{The PDF of the cosine of the angle between $\vec{a}$ and $\vec{v}$ of honeybee flight conditioned on flight radius $r$ for no wind conditions  while flying away from the hive (a) and while flying towards the hive (b). Also shown for windy conditions while flying away from the hive or flying against the wind (c) and flying towards the hive or flying in the same direction as the wind (d). The solid lines here are polynomial fits to the data shown with circles for better visualization.}
\label{away and towards r cond}
\end{figure}

The PDF of $\vec{a}\cdot \vec{v}/ \lvert \vec{a} \rvert \lvert \vec{v} \rvert$ can be conditioned on the radius of flight and is shown in Figure~\ref{away and towards r cond} for no wind conditions and windy conditions while flying away and towards the hive. We calculate the radius of honeybee flight by obtaining the centripetal acceleration and using the relation $a_{c}=v^2/r$. 
Honeybee flight is not significantly affected by radii smaller than the median radius where statistics are almost unchanged for $r< \text{M}(r)$ and $r< 1/2~\text{M}(r)$.
Thus, honeybees are most likely to fly in radii equal to or larger than that of the median radius. 
Here for no wind present, the median radius while flying away from the hive is 15.7~cm and while flying towards the hive is 13.8~cm. And for windy conditions the median radius while flying away from the hive is 21.9~cm and for flying towards the hive is 12.7~cm.
As can be seen from Figure~\ref{away and towards r cond} (a) and (c), the maxima at smaller values of the cosine at $0<\vec{a}\cdot \vec{v}/ \lvert \vec{a} \rvert \lvert \vec{v} \rvert<\sqrt{2}/2$ in no wind conditions, as compared to windy conditions, show that the honeybee is able to make smaller radius turns when wind is not present.
While, when the honeybee moves in windy conditions and flies against the wind, due to less flight control, it flies in larger radii. 
However, the median radius does not significantly change as the honeybee flies towards the hive in both no wind and windy conditions.

As the flight radius becomes larger and the flight path is closer to a straight line, the honeybee is almost always accelerating in no wind conditions while in windy conditions it is also braking and decelerating. 
During flight away from the hive in Figure~\ref{away and towards r cond} (a) and (c), we see a similar flight pattern for small radii as before.
In no wind conditions during acceleration, the honeybee acceleration is either translational or centripetal. While, in windy conditions during acceleration, the translational acceleration is suppressed and accelerations are mostly centripetal due to increased turning and maneuvering.
This is also observed in flight towards the hive in Figure~\ref{away and towards r cond} (b) and (d). Here, in no wind conditions, honeybee flight is evenly distributed between translational and centripetal accelerations during deceleration, while in windy conditions the honeybee has mostly centripetal acceleration during deceleration, as can be seen by the maxima at $-\sqrt{2}/2<\vec{a}\cdot \vec{v}/ \lvert \vec{a} \rvert \lvert \vec{v} \rvert<0$.
Overall, in windy conditions the honeybee flight mostly consists of turning and performing evasive maneuvers for both accelerating and decelerating events during flight away and towards the hive, while in no wind conditions part of the flight also consists of more straight flight with larger contributions to translational acceleration.

\section{Conclusions}

Statistically, the honeybee velocities, accelerations, and overall flight performance are not dependent on the turbulence intensity based on the experiments conducted in this study and the turbulence intensities generated.
To observe the influence of turbulence on honeybee flight we look at the angle between the velocity and acceleration vector, the cosine of this angle reveals interesting behavior in honeybee flight.
Form the cosine of this angle we see that honeybees in general spend the majority of their flight time accelerating, while deceleration events are sudden with extreme values of acceleration. This means that accelerations to top speeds take a longer time than decelerating to lower speeds, analogous to flight-crash events in turbulence~\cite{Xu2014}.

The cosine of the angle between $\vec{a}$ and $\vec{v}$ further shows that in calm conditions with no wind, a honeybee is either flying straight or making turns during flight.
However, as wind is introduced, straight flight is suppressed and flight is perturbed in a way that it  mostly consists of turning and performing evasive maneuvers.


Future work would also need to examine statistics in the reference frame of the honeybee to investigate the effects of body motion and rotation of the honeybee in turbulence. This can be achieved with more advanced cameras and imaging at a higher frame rate and pixel density, which was not reliably possible with the accessible GoPro system used in these experiments. The results presented here on how honeybees navigate turbulence along with future studies expanding on this work can help advance miniature robotics for more stable and effective flight resistant to damage and flight path deviations due to turbulent wind.

\section*{Supplementary information}
Further details of wind speed measurements while fans were off and when fans and active grid were on can be found in the Supplementary information.

\section*{Acknowledgments}

This work would not have been possible without the support of the technical staff at MPI-DS.
The authors would like to thank Sarah Romanowski and Maren S. M\"uller for taking care of the honeybee hives. We would also like to thank Artur Kubitzek, Andreas Renner, Andreas Kopp, Marcel Meyer, and Udo Schminke and his team at the Max Planck machine shop for help with making the turbulent active grid mobile and operational. We thank Marcel Schr\"oder for preparing the sonic anemometer for use.
We greatly appreciate Ramin Golestanian and Alexandre Mamane for insightful discussions.

\section*{Data and code availability}
Available upon reasonable request.


\section*{Author contributions}
BH, CK, EB planning and development of research project. BH, CK performing experiments. 
BH imaging and tracking. CK characterizing the turbulent flow. BH analysis of data and writing original draft. BH, CK, GB, EB discussions on the analysis, editing and preparation of final draft.

\section*{Funding}

We would like to thank the Max-Planck-Gesellschaft for support of this work.

\section*{Ethics statement}

According to the representative for animal experiments in basic research at the Max Planck Society, no ethical or legal statements are required for the experiments presented in this work.

\section*{Competing interests}
The authors declare no competing interests



\end{document}